\documentclass[amsmath,amssymb]{article}    

\usepackage{jheppub}
\usepackage{amsmath}
\usepackage{graphicx}
\usepackage{enumerate}
\usepackage{amsfonts}
\usepackage{graphicx}
\usepackage{enumerate}
\usepackage{setspace}
\usepackage[utf8]{inputenc}
\usepackage[T1]{fontenc}
\usepackage{hyperref}

\usepackage{bbm}
\usepackage{amssymb}
\usepackage{amsthm}
\usepackage{mathtools}
\usepackage{filecontents}
\usepackage{comment}
\usepackage{mathrsfs}

\makeatletter
\def\@fpheader{\relax}
\makeatother

\newcounter{parentsubequation}

\makeatother

\DeclareMathAlphabet{\mathbbold}{U}{bbold}{m}{n} 

\DeclareMathOperator{\Tr}{Tr}
\DeclareMathOperator{\diag}{diag}

\title{Expansions for semiclassical conformal blocks}

\author{Bruno Carneiro da Cunha, and}
\emailAdd{bruno.ccunha@ufpe.br}
\author{João Paulo Cavalcante}
\emailAdd{joaopaulocavalcante@hotmail.com.br}

\affiliation{Departamento de Física, Universidade Federal de
  Pernambuco, 50670-901, Recife, Brazil}

\abstract{We propose a relation the expansions of regular and irregular
  semiclassical conformal blocks at different branch points making use
  of the connection between the accessory parameters of the BPZ
  decoupling equations to the logarithm derivative of isomonodromic
  tau functions. We give support for these relations by considering
  two eigenvalue problems for the confluent Heun equations obtained
  from the linearized perturbation theory of black holes. We first
  derive the large frequency expansion of the spheroidal equations,
  and then compare numerically the excited quasi-normal mode spectrum
  for the Schwarzschild case obtained from the large frequency
  expansion to the one obtained from the low frequency expansion and
  with the literature, indicating that the relations hold generically
  in the complex modulus plane.}  


\begin{document}

\maketitle

\section{Introduction}

Conformal blocks are special functions that arise in the computation
of four point functions in conformal field theories. As such, they are
naturally dependent on the conformal dimensions of the operators
entering the correlation function. Less trivially, they also depend on
the internal dimension, sometimes called channel, as well as the
conformal modulus $t$ and the central charge of the Virasoro
algebra. Conformal blocks are then the building blocks of generic
correlation functions in CFTs.

The AGT correspondence \cite{Alday:2009aq} related the
conformal blocks to instanton partition functions of 4-dimensional
supersymmetric gauge theories. The latter had had a well-established
representation in terms of Nekrasov functions \cite{Nekrasov:2002qd},
and the following years saw a great surge of activity in understanding
how the correspondence works and applications of conformal blocks
outside of CFTs.

Let us mention two of those, with singular interest. First, we single
the formulation of the expansion of isomonodromic 
tau functions, originally defined in \cite{Jimbo:1981aa}, in terms of
$c=1$ conformal blocks $\mathscr{B}$. As a representative example, let
us present the \textit{Kyiv formula} for the Painlevé VI tau function,
first derived in \cite{Gamayun:2012ma}
\begin{equation}
  \begin{gathered}
    \tau_{VI}(\theta_k;\sigma,\eta;t)=\sum_{n\in\mathbb{Z}}e^{2\pi
      i n\eta}
    \mathscr{B}(\theta_k;\sigma+2n;t),\\
    \mathscr{B}(\theta_k;\sigma;t)=
    \mathscr{N}^{\theta_2}_{\theta_3,\sigma}
    \mathscr{N}^{\theta_1}_{\sigma,\theta_0}
    t^{\frac{1}{4}(\sigma^2-\theta_0^2-\theta_1^2)}(1-t)^{\frac{1}{2}\theta_1\theta_2}
    \sum_{\lambda,\mu\in\mathbb{Y}}\mathscr{B}_{\lambda,\mu}(\theta_k,\sigma)
    t^{|\lambda|+|\mu|},
  \end{gathered}
  \label{eq:taunekrasov6}
\end{equation}
where the sum on $\mathbb{Y}$ are over pairs $(\lambda,\mu)$ of Young
diagrams, and 
\begin{multline}
  \mathscr{B}_{\lambda,\mu}(\theta_k,\sigma) =
  \prod_{(i,j)\in\lambda}\frac{((\theta_1+\sigma+2(i-j))^2-\theta_0^2)
    ((\theta_2+\sigma+2(i-j))^2-\theta_3^2)}{16h_\lambda^2(i,j)
    (\lambda'_j-i+\mu_i-j+1+\sigma)^2}\times\\
  \prod_{(i,j)\in\mu}\frac{((\theta_1-\sigma+2(i-j))^2-\theta_0^2)
    ((\theta_2-\sigma+2(i-j))^2-\theta_3^2)}{16h_\mu^2(i,j)
    (\mu'_j-i+\lambda_i-j+1-\sigma)^2},
\end{multline}
and
\begin{equation}
  \mathscr{N}^{\phi_2}_{\phi_3,\phi_1}=
  \frac{\prod_{\epsilon=\pm}
    G(1+\frac{1}{2}(\phi_3+\epsilon(\phi_1+\phi_2)))
    G(1-\frac{1}{2}(\phi_3+\epsilon(\phi_1-\phi_2)))}{
    G(1-\phi_1)G(1-\phi_2)G(1+\phi_3)},
\end{equation}
where $G(z)$ is the Barnes' function. Similar expansions are also
available for Painlevé V and III \cite{Gamayun:2013auu}, which can be
seen as confluent limits of \eqref{eq:taunekrasov6}, making use of
irregular conformal blocks \cite{Gaiotto:2009ma}, as explained in
\cite{Nagoya:2015cja,Lisovyy:2018mnj}. 

More recently, semiclassical conformal blocks have also been proposed
as solutions to the connection problem of solutions to the Heun
equation \cite{Bonelli:2022ten}, in the \textit{Trieste formula}
\cite{Lisovyy:2022flm}
\begin{equation}
  \mathscr{C}_{01}=\frac{\Gamma(1-\theta_0)\Gamma(\theta_2)}{
    \Gamma(\tfrac{1}{2}(1+\theta_2-\theta_0+\theta_3))
    \Gamma(\tfrac{1}{2}(1+\theta_2-\theta_0-\theta_3))}
  \exp\left[
      \frac{\partial\mathscr{W}}{\partial\theta_2}-
      \frac{\partial\mathscr{W}}{\partial\theta_0}\right],
    \label{eq:connectionheun}
\end{equation}
which makes extensive use of the relation between semiclassical
conformal blocks $\mathscr{W}$ and accessory parameters
\cite{Belavin:1984vu,Zamolodchikov:1987aa}. Here $\mathscr{C}_{01}$ is
the connection coefficients between Frobenius solutions of the Heun
equation constructed at $z=0$ and $z=1$. Similar relations involving
accessory parameters for the same equation have been derived using
different methods in \cite{Lisovyy:2021bkm}. 

Surprisingly, both $c=1$ and semiclassical conformal blocks are
related through the accessory parameter of Heun equations. As
demonstrated in a variety of manners
\cite{Novaes:2014lha,Anselmo:2018,Lencses:2017dgf,CarneirodaCunha:2019tia,Bershtein:2021uts},
we have schematically that the accessory parameter is given by
\begin{equation}
  c = \frac{\partial \mathscr{W}}{\partial t} =
  \frac{\partial}{\partial t}\log\tau(t),
\end{equation}
where the logarithmic derivative is defined on the zero locus of an
associated tau function, as made explicit in the references and revised
in \eqref{eq:c0tau6} and \eqref{eq:irregconfblktau5} below. In its
latest development, this relation hints at an equivalence between the
semiclassical conformal block and the logarithm of the tau function,
computed at its zero locus. This seems to be a consequence of the
blow-up formulas in gauge theory \cite{Nakajima:2003aa}, but we are
unaware of a treatment that covers the conformal blocks entering the
Painlevé VI and V tau functions. 

On the other hand, if one takes the relation as true, one can use
known braid transformations of the Painlevé tau functions to derive
expansions for the conformal blocks at different singular
points. Usually, these singular points are related to the collision of
vertex operators \cite{Zamolodchikov:1987aa}, and fusion rules can be
used to relate the expansions. For the irregular case, these are not
known explicitly.

In their own right, accessory parameters of Fuchsian differential
equations are a central problem in mathematical physics in applied
mathematics with a multitude of applications. As examples, we single
black hole quasi-normal mode calculations
\cite{daCunha:2015ana,Barragan-Amado:2018pxh,Novaes:2018fry,Aminov:2020yma,daCunha:2021jkm},
constructive conformal mapping
\cite{Anselmo:2018,Anselmo:2020aa,CarneirodaCunha:2022aa}, 
the Rabi model in quantum optics \cite{daCunha:2015npa} and generic
remarks on the accessory parameter problem in
\cite{Piatek:2014lma,Piatek:2017fyn}. In all of these, the role of the
monodromy parameters in formulating the expansion of the accessory
parameter is stressed.

In this paper, we explore the symplectic structure behind the
isomonodromic tau functions to relate expansions of the conformal
block at different singular points, both for the regular and irregular
rank 1 cases. While the structure is quite symmetric in the regular
case \eqref{eq:w0around0}, \eqref{eq:w0around1} and
\eqref{eq:w0aroundinfty}, as it is expected, the structure is less so
in the irregular case \eqref{eq:w0csmallt}, \eqref{eq:w0clarget}. We
also present a continued fraction algorithm to compute these conformal
blocks and use the relation between conformal blocks and accessory
parameter of the confluent Heun equation to study spheroidal
eigenvalues and quasi-normal modes for the Kerr black hole. We stress
the importance of better understanding the global analytical property
of conformal blocks, which are relevant not only to the applications
listed above, but also in their characterization as special functions
in their own right.

\section{Conformal blocks: Ward Identities}
\label{sec:ccb}

Let us consider chiral Liouville Field Theory \cite{Belavin:1984vu}
with central charge $C=1+6Q^2$, $Q=b^{-1}+b$. Let $V_{\alpha}(z)$ be
the local field operator corresponding to the classical field
$e^{(Q+\alpha)\varphi(z)}$. The Hilbert space of this theory
decomposes into Verma modules, generated from the action
of the Virasoro generators on a primary state resulting from the action of
$V_{\alpha}(0)$ on the $sl_2$-invariant state $|0\rangle$. The primary
state $|\alpha\rangle$ thus generated has conformal weight
$\Delta(\alpha)=(Q^2-\alpha^2)/4$. With the state-operator
correspondence, we can identify correlation functions with matrix
elements of a product of primary operators between two primary
states. These correlation functions admit the partial wave
decomposition
\begin{equation}
  \begin{aligned}
  \langle \alpha_{n-1} | V_{(1,2)}(z)V_{\alpha_{n-2}}(1) \ldots
  V_{\alpha_1}(z_1)|\alpha_0\rangle
  & =\sum_{\{\beta_k\}}\langle
  \alpha_{n-1}|V_{(1,2)}(z)V_{\alpha_{n-2}}(1)\Pi_{\beta_{n-2}}\ldots
  \Pi_{\beta_1} V_{\alpha_1}(z_1)|\alpha_0\rangle, \\
  & = \sum_{\{\beta_k\}}\mathscr{C}(\alpha_{n-1},\alpha_{n-2},\beta_{n-2})
  \ldots \mathscr{C}(\beta_1,\alpha_1,\alpha_0)
    \mathscr{G}_k(z,z_k;\alpha_k,\beta_k),
  \end{aligned}
  \label{eq:confblk}
\end{equation}
where $\Pi_{\beta_k}$ is the projection operator on the Verma module
built from $|\beta_k\rangle$ and $\mathscr{C}(\alpha_i,\alpha_j,\alpha_k)$ is the
Liouville structure constant. The function
$\mathscr{G}_{k}(z,z_k;\alpha_k,\beta_k)$ defined in the left 
hand side of \eqref{eq:confblk} are, up to normalization, the Virasoro
conformal blocks. 

It has long been known that for special values of
$\alpha=\alpha_{n,m}=-nb^{-1}-mb$, $n,m=1,2,\ldots$, the
associated Verma module corresponds to degenerate representations of
the Virasoro algebra. For instance, for $\alpha_{(1,2)}=-b^{-1}-2b$, there
is a null state 
\begin{equation}
  \frac{1}{b^2}\frac{\partial^2}{\partial z^2} V_{(1,2)}(z)
  +:T(z)V_{(1,2)}(z):=0,
  \label{eq:level2null}
\end{equation}
which decouples from correlation functions with local operators. This
in turn means that regular conformal blocks involving $V_{(1,2)}(z)$ will
satisfy second order differential equations of the Fuchsian
type. After using the global conformal group to fix the position of
three of those primary operators to $z=0,1,$ and $\infty$, these
equations will have the form 
\begin{multline}
  \left(\frac{1}{b^2}\frac{\partial^2}{\partial z^2}-
    \left(\frac{1}{z}+\frac{1}{z-1}\right)\frac{\partial}{\partial z}+
    \frac{\Delta_0}{z^2}+\frac{\Delta_{n-2}}{(z-1)^2}+
    \sum_{k=1}^{n-3}\frac{\Delta_k}{(z-z_k)^2}
    +\right. \\ \left.
    \frac{\Delta_{n-1}-\Delta_{(1,2)}-\sum_{k=0}^{n-2}\Delta_k}{z(z-1)}
    +\sum_{k=1}^{n-3}
    \frac{z_k(z_k-1)}{z(z-1)(z-z_k)}\frac{\partial}{\partial z_k}
  \right)\mathscr{G}_{k}(z,z_k;\alpha_k,\beta_k)=0,
  \label{eq:bpzregular}
\end{multline}
where $z_0=0$ and $z_{n-2}=1$.

In the semiclassical limit of Liouville Field Theory, defined as
\begin{equation}
  \alpha_k= \frac{\theta_k}{b},
  \qquad \beta_k= \frac{\sigma_k}{b}, \qquad
  b\rightarrow 0, 
\end{equation}
the conformal blocks are expected to exponentiate
\cite{Zamolodchikov:1987aa,Teschner:2010je,Litvinov:2013sxa}:
\begin{equation}
  \mathscr{G}_k(z,z_k;\theta_k,\sigma_k) =
  \psi(z,z_k;\theta_k,\sigma_k)
  \exp\left[\frac{1}{b^2}\mathscr{W}(z_k;\theta_k,\sigma_k)\right]
  (1+{\cal O}(b^2)).
\end{equation}
The wave functions $\psi(z,z_k;\theta_k,\sigma_k)$ now satisfy an
ordinary differential equation with regular singular points similar to
\eqref{eq:bpzregular}, where the accessory parameters
\begin{equation}
  c_k(\sigma_k,z_k)=
  \frac{\partial}{\partial z_k}\mathscr{W}(z_k;\theta_k,\sigma_k),
  \label{eq:accparam6}
\end{equation}
with the $\theta_k$ dependence omitted, replace the derivatives with
respect to $z_k$ in the BPZ decoupling 
equation \eqref{eq:bpzregular},
\begin{equation}
  \left(\frac{\partial^2}{\partial z^2}+
    \sum_{k=0}^{n-2}\frac{\delta_k}{(z-z_k)^2}+
    \frac{\delta_{n-1}-\sum_{k=0}^{n-2}\delta_k}{z(z-1)}
    +\sum_{k=1}^{n-3}\frac{z_k(z_k-1)c_k}{z(z-1)(z-z_k)}
  \right)\psi(z,z_k;\theta_k,\sigma_k)=0,
  \label{eq:bpzregsc}
\end{equation}
with $\delta_k=\lim b^2\Delta_k=(1-\theta_k^2)/4$.

The equation \eqref{eq:bpzregsc} is the most general second order
Fuchsian differential equation with $n$ singular points. The relation
between accessory parameters $c_k$ and semi-classical Virasoro
conformal blocks is known \cite{Zamolodchikov:1987aa}, and it is
deeply tied to the monodromy problem of finding $c_k$ and $z_k$ in
\eqref{eq:bpzregsc} such that $\psi(z,z_k;\theta_k,\sigma_k)$ has
prescribed monodromy properties.

The solution to the general monodromy problem was outlined in
\cite{Litvinov:2013sxa} and made explicit in
\cite{CarneirodaCunha:2022aa} in terms of logarithm derivative of the
isomonodromic Jimbo-Miwa-Ueno tau function, introduced in
\cite{Jimbo:1981aa}, and whose generic expansion was given in
\cite{Gavrylenko:2016zlf}. The relation between these functions and
the semiclassical conformal blocks is deep, only really understood
from the AGT correspondence in gauge field language
\cite{Nakajima:2003aa}. Our purpose in mentioning the monodromy
problem is more direct, having to do with a global definition of the
semiclassical conformal block $\mathscr{W}$. The monodromy property
stems from the OPE between $V_{(1,2)}(z)$ and a generic primary
\begin{multline}
  V_{(1,2)}(z)V_{\alpha}(w)=A_+(z-w)^{b(Q-\alpha)/2}
  (V_{\alpha+b}(w)+{\cal O}(z-w))+\\
  A_-(z-w)^{b(Q+\alpha)/2}(V_{\alpha-b}(w)+{\cal O}(z-w)),
  \label{eq:nullOPE}
\end{multline}
where the structure constants $A_\pm$ can be computed from the DOZZ
formula, see for instance \cite{Teschner:1995yf}. Since both expansions in
\eqref{eq:nullOPE} are analytic for $z$ sufficiently close to $w$, one
can read the monodromy properties of $\mathscr{G}_k$ from the sets of
$\alpha_k$ and $\beta_k$. We propose that this suffices to define the
semiclassical conformal block $\mathscr{W}$ globally with respect to
the $z_k$ variables.

Let us illustrate this construction for 4 regular singular ponts,
where \eqref{eq:bpzregsc} is known as the Heun equation
\cite{Ronveaux:1995}. Let us call the conformal modulus and the
accessory parameter $z_1=t$ and $c_1=c$, and $\mathscr{W}$ is a
function of $t$, $\theta_k$ and a single $\sigma$, which parametrizes the
momentum of the intermediate channel. There are a variety of methods
of computing $\mathscr{W}$ as a perturbative series in $t$ in this
case, from Zamolodchikov's recursion formula
\cite{Zamolodchikov:1987aa}, the Nekrasov-Shatashvili limit of the
Nekrasov function \cite{Nekrasov:2002qd,Nekrasov:2009rc}. More recently
\cite{Novaes:2014lha,Lencses:2017dgf,Anselmo:2018,Bershtein:2021uts},
the following equations involving \eqref{eq:taunekrasov6} and $c_k$: 
\begin{equation}
  c(\sigma,t)=\frac{\partial}{\partial t}
  \log\tau_{VI}(\theta^{-}_k;\sigma,\eta;t)+
  \frac{\theta_1-1}{2t}+\frac{\theta_1-1}{2(t-1)},\qquad
  \tau_{VI}(\theta_k;\sigma+1,\eta;t)=0,
  \label{eq:c0tau6}
\end{equation}
where $\theta^-_k=\{\theta_0,\theta_1-1,\theta_2,\theta_3+1\}$. 
These equations hint at the relation between $\mathscr{W}$ and the
logarithm derivative of the tau function, at least on its zero locus. The second
equation has many different representations \cite{Okamoto:1986aa} and
it is sometimes called Toda equation. The introduction of the $\eta$
parameter seems like an unnecessary complication, but it helps to give
the Toda equation a geometrical interpretation. Borrowing from the
isomonodromy construction of the tau function, we will think of a
configuration space parametrized by $\sigma,\eta$ and $t$. In turn,
the $\sigma$ and $\eta$ parameters are (local) coordinates on the
manifold of monodromy data. The zero locus condition above then
defines a Lagrangian submanifold of the isomonodromic flow in a manner
independent of the local coordinates chosen to parametrize the
manifold of monodromy data. The introduction of $\eta$ as an
independent variable allows us to take \eqref{eq:c0tau6} as a 
global definition for $c$, and therefore of $\mathscr{W}$. 

The tau function \eqref{eq:taunekrasov6} has the Painlevé property
\cite{Miwa:1981aa}: apart from the fixed branch points at
$t=0,1$ and $\infty$, it is an analytic function of $t$. Given the
definition of the tau function in terms of isomonodromy flow, the
transformation laws can be readily established \cite{Jimbo:1982aa}
\begin{equation}
  \begin{aligned}
    \tau_{VI}(\theta_k;\sigma,\eta;t) &
    = {\cal N}_{01}
    \tau_{VI}(\tilde{\theta}_k;\tilde{\sigma},\tilde{\eta};1-t),\qquad
    &\tilde{\theta}_k=\{\theta_2,\theta_1,\theta_0,\theta_3\},\\ 
    & = {\cal N}_{0\infty}
    \tau_{VI}(\bar{\theta}_k;\bar{\sigma},\bar{\eta};1/t),\qquad
    &\bar{\theta}_k=\{\theta_3,\theta_1,\theta_2,\theta_0\},
  \end{aligned}
  \label{eq:taubraidtrans}
\end{equation}
where the connection coefficients at $t$ independent and were computed in
\cite{Iorgov:2013uoa}. The new monodromy variables
$\tilde{\sigma},\tilde{\eta}$ and $\bar{\sigma}$ and $\bar{\eta}$ are
related to the monodromy of $C=1$ conformal blocks, but here we can
reinterpret them in terms of semiclassical fusion rules. At any rate,
we note that equations \eqref{eq:c0tau6} are invariant under these
transformations and then one can use the transformations
\eqref{eq:taubraidtrans} to derive expansions for the
accessory parameters -- and hence for $\mathscr{W}$ -- at different
branching points once the transformation laws for the monodromy
parameters are obtained. We will list the relations between the
monodromy parameters below.

Let us now return to \eqref{eq:nullOPE}. The OPE between the
degenerate field and one of the primaries will result in two series
with different local monodromies. These series can be identified, up
to normalization, with local Frobenius solutions of the Heun equation
$\psi_{k,\pm}(z)$ at $z_k$. One can in principle work out which particular
solution of \eqref{eq:bpzregsc} is realized by the conformal block
involving $V_{(1,2)}(z)$, but that will not be necessary to our
purposes.

Let us call $\mathbf{M}_k$ the monodromy matrix associated to the
analytic continuation of a generic pair of solutions $\psi_{\pm}(z)$
around $z_k$. If one picks the pair to be the Frobenius solutions
built on $z_k$, then the monodromy matrix will be diagonal
$\mathbf{M}_k = \diag(-e^{\pi i\theta_k},-e^{-\pi i\theta_k})$, but
of course this need not be the case if one chooses a monodromy path
around a different singular point, or indeed a different pair of solutions. 

To associate the intermediate momentum parameter $\sigma$ with
monodromy, let us recall that, for the small $t$ expansion of
$\mathscr{W}$, $\beta=\sigma b^{-1}$ comes from the OPE between
$V_{\alpha_0}(0)$ and $V_{\alpha_1}(t)$. In the radial quantization
picture, the substitution makes sense for $|z|>|t|$, so then the
monodromy related to $\sigma$ is associated to a path encompassing
both $t$ and $0$. In terms of monodromy matrices,
\begin{equation}
  2\cos\pi\sigma = -\Tr \mathbf{M}_{1}\mathbf{M}_0.
\end{equation}
By the same token, expansions at $t=1$ and $t=\infty$ are related
to the OPEs between $V_{\alpha_1}(t)$ and $V_{\alpha_2}(1)$ and
$V_{\alpha_3}(\infty)$, respectively, so we conclude
\begin{gather}
  2\cos\pi\tilde{\sigma}=-\Tr\mathbf{M}_{2}\mathbf{M}_{1},\\
  2\cos\pi\bar{\sigma}=-\Tr\mathbf{M}_{3}\mathbf{M}_{1}.
  \label{eq:tildebarsigma}
\end{gather}

The $\eta$ coordinate is the canonical conjugate to $\sigma$ in the
symplectic structure of the moduli space of flat meromorphic
connections on a 4-punctured Riemann surface, as explained in
\cite{Nekrasov:2011bc,Iorgov:2014vla}. We will use the more direct
explanation from \cite{Its:2016jkt}. Consider a pair of Floquet
solutions to \eqref{eq:bpzregsc}
\begin{equation}
  \psi_{F,\pm}(z,t;\theta_k,\sigma) =
  z^{\frac{1}{2}(1+\theta_1\pm\sigma)}(z-t)^{\frac{1}{2}(1-\theta_1)}
  (z-1)^{\frac{1}{2}(1-\theta_2)}\sum_{n\in\mathbb{Z}}a_nz^n,
  \label{eq:floquetheun}
\end{equation}
where we assume that the Laurent series converges to an analytic
function in an annulus containing $0$ and $t$, but not $1$. In this
basis, the monodromy matrix around $0$ and $t$ is diagonal, so we
can write
\begin{equation}
  \mathbf{M}_{1}\mathbf{M}_{0}=-e^{\pi i\sigma
    \left(
      \begin{smallmatrix} 1 & 0 \\ 0 & -1 \end{smallmatrix}
  \right)} =
  \mathbf{M}_2^{-1}\mathbf{M}_{3}^{-1},
\end{equation}
where the last equality is the result of the contractibility of the
loop circling around all four punctures. The problem of finding two 
unimodular matrices with known traces that multiply to a diagonal
matrix can be solved algebraically (see Eqns. (3.32) and (3.33) in
\cite{Its:2016jkt}). The solution, however, is not unique: conjugation
by a diagonal matrix ($\mathbf{M}_k\rightarrow
s^{\sigma_3}\mathbf{M}_ks^{-\sigma_3}$, with $\sigma_3=\diag(1,-1)$)
leaves the product -- itself a diagonal matrix -- invariant. 

This ambiguity introduces two parameters $s_i$ and $s_e$ to the
explicit parametrization of the $\mathbf{M}_k$. Given a solution to
the equation
$\hat{\mathbf{M}}_{B}(\phi_1,\phi_2,\phi_3)
\hat{\mathbf{M}}_{A}(\phi_1,\phi_2,\phi_3)
=-\exp(\pi i\phi_3\sigma_3)$, we can then define
\begin{equation}
  \begin{gathered}
    \mathbf{M}_0=s_i^{\sigma_3}\hat{\mathbf{M}}_A(\theta_0,\theta_1,\sigma)
    s_i^{-\sigma_3},\qquad
    \mathbf{M}_1=s_i^{\sigma_3}\hat{\mathbf{M}}_B(\theta_0,\theta_1,\sigma)
    s_i^{-\sigma_3},\\
    \mathbf{M}_2=s_e^{\sigma_3}\hat{\mathbf{M}}_A(\theta_2,\theta_3,-\sigma)
    s_e^{-\sigma_3},\qquad
    \mathbf{M}_3=s_e^{\sigma_3}\hat{\mathbf{M}}_B(\theta_2,\theta_3,-\sigma)
    s_e^{-\sigma_3}.
    \label{eq:explicitems}
  \end{gathered}
\end{equation}
From this explicit representation we can see that only the ratio
$s_e/s_i$ is invariant by change of basis. We thus define $\eta$ as 
\begin{equation}
  e^{\pi i\eta}=\frac{s_i}{s_e},
  \label{eq:etamonodromy}
\end{equation}
With \eqref{eq:explicitems} one can compute $\tilde{\sigma}$ and
$\bar{\sigma}$ from \eqref{eq:tildebarsigma}, arriving at
\begin{multline}
  \sin^2\pi\sigma (\cos\pi\tilde{\sigma}+\cos\pi(\theta_1+\theta_2))
  =\\
   2\cos\tfrac{\pi}{2}(\sigma-\theta_2+\theta_3)
  \cos\tfrac{\pi}{2}(\sigma-\theta_2-\theta_3)
  \cos\tfrac{\pi}{2}(\sigma-\theta_1+\theta_0)
  \cos\tfrac{\pi}{2}(\sigma-\theta_1-\theta_0)(e^{2\pi i\eta}-1)+\\
  2\cos\tfrac{\pi}{2}(\sigma+\theta_2+\theta_3)
  \cos\tfrac{\pi}{2}(\sigma+\theta_2-\theta_3)
  \cos\tfrac{\pi}{2}(\sigma+\theta_1+\theta_0)
  \cos\tfrac{\pi}{2}(\sigma+\theta_1-\theta_0)(e^{-2\pi i \eta}-1).
  \label{eq:sigmat1}
\end{multline}
and
\begin{multline}
  \sin^2\pi\sigma (\cos\pi\bar{\sigma}
  +e^{\pi i(\theta_1+\theta_2)}\cos\pi\sigma+e^{\pi
    i\theta_1}\cos\pi\theta_3 +e^{\pi i\theta_2}\cos\pi\theta_0)
  =\\
    2\cos\tfrac{\pi}{2}(\sigma-\theta_2+\theta_3)
  \cos\tfrac{\pi}{2}(\sigma-\theta_2-\theta_3)
  \cos\tfrac{\pi}{2}(\sigma-\theta_1+\theta_0)
  \cos\tfrac{\pi}{2}(\sigma-\theta_1-\theta_0)
  e^{-\pi i\sigma}(e^{2\pi i \eta}-1)+\\
  2\cos\tfrac{\pi}{2}(\sigma+\theta_2+\theta_3)
  \cos\tfrac{\pi}{2}(\sigma+\theta_2-\theta_3)
  \cos\tfrac{\pi}{2}(\sigma+\theta_1+\theta_0)
  \cos\tfrac{\pi}{2}(\sigma+\theta_1-\theta_0)
  e^{\pi i\sigma}(e^{-2\pi i \eta}-1).
  \label{eq:sigma01}
\end{multline}
We note that the braid transformations \eqref{eq:taubraidtrans}
\begin{equation}
  \begin{gathered}
    \sigma\rightarrow\tilde{\sigma},\qquad
    \theta_0\leftrightarrow\theta_2, \qquad\text{and}\qquad
    \sigma\rightarrow\bar{\sigma},\qquad
    \theta_0\leftrightarrow\theta_3,
  \end{gathered}
\end{equation}
allows us to define canonically conjugate variables $\tilde{\eta}$ and
$\bar{\eta}$ analogously to \eqref{eq:etamonodromy}.

The symplectic structure is illustrated by defining the Fricke-Jimbo
polynomial
\begin{multline}
  W(p_{kl}) = p_0p_{1}p_2p_{3}-p_{01}p_{12}p_{02}+
  (p_0p_{1}+p_2p_{3})p_{01}+
  (p_{1}p_2+p_0p_{3})p_{12}+\\
  +(p_0p_2+p_{1}p_{3})p_{02}
  +p_{01}^2+p_{12}^2+p_{02}^2+p_0^2+p_{1}^2+p_2^2+p_3^2-4,
\end{multline}
where $p_k=-2\cos\pi\theta_k$ and $p_{kl}=-2\cos\pi\sigma_{kl}$, with
$\sigma_{01}=\sigma$, $\sigma_{12}=\tilde{\sigma}$ and
$\sigma_{02}=\bar{\sigma}$ are trace coordinates. It can be shown that
for any monodromy matrices $\mathbf{M}_k$, $W(p_{kl})=0$, so only two
of the $p_{kl}$ are independent. This space can be furnished with a
symplectic structure \cite{Goldman:2009aa}:
\begin{equation}
  \{ p_{\mu},p_{\nu} \} = -\frac{\partial W}{\partial p_{\rho}}
\end{equation}
where $\mu,\nu,\rho$ cycle around the three pairs $01,12$, and
$02$. It is now a straightforward exercise (see Lemma 3.12 in
\cite{Its:2016jkt}) to show that $\sigma$ and $\eta$ are canonical
Darboux coordinates of the monodromy parameter space,
$\{\eta,\sigma\}=1$. Again in complete analogy we have
$\{\tilde{\eta},\tilde{\sigma}\}=\{\bar{\eta},\bar{\sigma}\}=1$ as a
result of the braid transformations \eqref{eq:taubraidtrans}. 

Being the canonical conjugate coordinate to $\sigma$ means that $\eta$
is not in fact independent, as hinted from \eqref{eq:c0tau6}. The key
for computing it from knowledge of $\sigma$ and $t$ is the zero of
$\tau_{VI}$. The structure of $\tau_{VI}$ is such that it can be seen
as an analytic function of $t$ and meromorphic in $\chi =
t^{\sigma}e^{2\pi i\eta}$ \cite{Jimbo:1982aa}. We can then formally
invert the expansion $\tau_{VI}=0$ and write a series
\begin{equation}
  \chi = \Theta(\sigma,\theta_2,\theta_3)
  \Theta(\sigma,\theta_1,\theta_0)(1+\chi_1t+\chi_2t^2+\ldots),
\end{equation}
and compute $\chi_k$ from the conformal blocks, see
\cite{Barragan-Amado:2018pxh} for the first coefficients. The function
\begin{equation}
  \Theta(\sigma,\phi_1,\phi_2) =
  \frac{\Gamma(1+\sigma)}{\Gamma(1-\sigma)}
  \frac{\Gamma(\tfrac{1}{2}(1+\phi_1+\phi_2-\sigma))
    \Gamma(\tfrac{1}{2}(1+\phi_1-\phi_2-\sigma))}{
    \Gamma(\tfrac{1}{2}(1+\phi_1+\phi_2+\sigma))
    \Gamma(\tfrac{1}{2}(1+\phi_1-\phi_2+\sigma))},
  \label{eq:theta6}
\end{equation}
comes about from the ratio of the Barnes' $G$-functions in the
$\mathscr{N}$ coefficients in \eqref{eq:taunekrasov6}. 

Since $\eta$ is canonically conjugated to $\sigma$, we expect
that its $t$ dependence to be given by semiclassical conformal block
through its $\sigma$ derivative.  Given the relation
\eqref{eq:accparam6} and the $t$-independent term in the definition of
$\eta$, we can conjecture
\begin{equation}
  \eta(\sigma,t) = \frac{1}{2\pi i}\log \Theta(\sigma,\theta_1,\theta_0)
  \Theta(\sigma,\theta_2,\theta_3)+
  \frac{1}{\pi i}\frac{\partial}{\partial
    \sigma}\int^tdt'\,c(\sigma,t)
  =\frac{1}{\pi i}\frac{\partial \mathscr{W}_s}{\partial \sigma},
  \label{eq:etaasw6}
\end{equation}
where we can fix the term independent of $t$ by comparing with the
expansion for $\eta(\sigma,t)$ obtained from the condition of zero of the tau
function. With this in mind, now define the small $t$ expansion of the
conformal block $\mathscr{W}_s$ with the $t$-independent term as 
\begin{equation}
  \Omega(\theta_k,\sigma) = \frac{1}{2}\int^\sigma d\sigma'
  \log
  \Theta(\sigma',\theta_1,\theta_0)\Theta(\sigma',\theta_2,\theta_3),
\end{equation}
and we note that this integral can be expressed in terms of Barnes'
$G$-functions -- see, for instance, (5.17.4) in 
\href{https://dlmf.nist.gov/5.17}{NIST's} Digital Library for
Mathematical Functions.  

The expansion for $c(\sigma,t)$ can be obtained efficiently from a
continued fraction method
\cite{daCunha:2021jkm,Lisovyy:2021bkm,Lisovyy:2022flm}.  We 
take the expansion \eqref{eq:floquetheun}, and by substituting in
Heun's equation, we obtain the 3-term recurrence relation
\begin{equation}
  A_na_{n-1}-(B_n+t\,C_n)a_n+t\,D_na_{n+1}=0,
  \label{eq:recurrenceeqnheun}
\end{equation}
with
\begin{equation}
  \begin{gathered}
    A_n=(\sigma+2n-1-\theta_2)^2-\theta_3^2,\qquad
    B_n=(\sigma+2n)^2-\theta_0^2-\theta_1^2+1,\\
    C_n=(\sigma+2n+\theta_1-\theta_2)^2+\theta_1^2-\theta_3^2-1
    -4(t-1)c,\qquad
    D_n=(\sigma+2n+1+\theta_1)^2-\theta_0^2.
  \end{gathered}
\end{equation}

The strategy to solving \eqref{eq:recurrenceeqnheun} is as
follows. First divide it by $a_{n}$ to find
\begin{equation}
  v_{n} = \frac{a_{n}}{a_{n+1}}=\frac{t\,D_n}{B_n+t\,C_n-A_nv_{n-1}},
\end{equation}
which allows to compute $v_{0}$ recursively from the knowledge of
$v_{n}$ for $n>0$. Given the factors of $t$, this expression can be
used to give a relation between $c(t)$ and $\sigma$ valid for small
$t$. Following the same procedure for $u_n = \frac{a_{n-1}}{a_n}$,
we consider \eqref{eq:recurrenceeqnheun} for $n=0$:
\begin{multline}
  \cfrac{t\,{A}_0{D}_{-1}}{{B}_{-1}+t\,{C}_{-1}
    -\cfrac{t\,{A}_{-1}{D}_{-2}}{{B}_{-2}+t\,{C}_{-2}
      -\cfrac{t\,{A}_{-2}{D}_{-3}}{{B}_{-3}+\ldots}}}+
  \cfrac{t\,{D}_0{A}_1}{{B}_1+t\,{C}_1-
    \cfrac{t\,{D}_1{A}_2}{{B}_2+t\,{C}_2-
      \cfrac{t\,{D}_2{A}_3}{{B}_3+\ldots}}}
  =  B_0+t\,{C}_0.
  \label{eq:contfracheun}
\end{multline}
Assuming a small $t$ expansion for $c(\sigma,t)$,
\begin{equation}
  c(\sigma,t) = c_{-1}(\sigma)t^{-1}+c_{0}(\sigma)+c_{1}(\sigma)t
  +\ldots+c_{n}(\sigma)t^n+\ldots
  \label{eq:c0around0}
\end{equation}
we can truncate \eqref{eq:contfracheun} to the $n$-th approximant to
compute $c_{n}$. The first terms are
\begin{gather}
  c_{-1}(\sigma)
  =\delta_\sigma-\delta_0-\delta_1,\\
  c_{0}(\sigma)=\frac{(\delta_\sigma-\delta_0+\delta_1)
    (\delta_\sigma-\delta_3+\delta_2)}{2\delta_\sigma},\\
 \begin{split}
  c_{1}(\sigma) =& \frac{(\delta_\sigma-\delta_0+\delta_1)
    (\delta_\sigma-\delta_3+\delta_2)}{2\delta_\sigma}
  -\frac{((\delta_0-\delta_1)^2-\delta_\sigma^2)
    ((\delta_2-\delta_3)^2-\delta_\sigma^2)}{8\delta_\sigma^3}\\
  & +\frac{(\delta_\sigma^2+2\delta_\sigma(\delta_0+\delta_1)-
    3(\delta_0-\delta_1)^2)(\delta_\sigma^2+2\delta_\sigma
    (\delta_3+\delta_2)-3(\delta_3-\delta_2)^2)}{32
    \delta_\sigma^2(\delta_\sigma+\tfrac{3}{4})},
  \end{split}
\end{gather}
where $\delta_k=(1-\theta_k^2)/4$ and
$\delta_\sigma=(1-\sigma^2)/4$. These agree with the expansion found
in the literature, as, for instance in \cite{Litvinov:2013sxa}. Given
the condition \eqref{eq:etaasw6}, we then see that the small $t$
expansion of the conformal block $\mathscr{W}_s(\sigma,t)$ to be the
primitive of $c(\sigma,t)$, with the $t$-independent term given by
$\Omega(\sigma)$: 
\begin{equation}
  \mathscr{W}_s(\sigma,t)=c_{-1}(\sigma)\log
  t+\Omega(\sigma)+c_{0}(\sigma)t+\frac{1}{2}c_{1}(\sigma) t^2+
  \ldots+\frac{1}{n}c_{n-1}(\sigma)t^n+\ldots.
  \label{eq:w0around0}
\end{equation}
By the same token, given the intermediate channel parameters
$\tilde{\sigma}$ and $\bar{\sigma}$, we can define the expansion near
$t=1$:
\begin{equation}
  \mathscr{W}_t(\tilde{\sigma},t)=
  \tilde{c}_{-1}(\tilde{\sigma})\log(1-t)+\Omega(\tilde{\sigma})+
  \tilde{c}_{0}(\tilde{\sigma})(1-t)+\ldots
  +\frac{1}{n}\tilde{c}_{n-1}(\tilde{\sigma})(1-t)^n+\ldots
  \label{eq:w0around1}
\end{equation}
where $\tilde{c}_{n}(\tilde{\sigma})$ is obtained from
$c_{n}(\sigma)$ by the interchange $\delta_0\leftrightarrow\delta_2$
and $\sigma\rightarrow\tilde{\sigma}$. And we have the expansion for
large $t$ as
\begin{equation}
  \mathscr{W}_u(\bar{\sigma},t)=
  -\bar{c}_{-1}(\bar{\sigma})\log t+\Omega(\bar{\sigma})
  +\bar{c}_{0}(\bar{\sigma})t^{-1}+\ldots
  +\frac{1}{n}\tilde{c}_{n-1}(\bar{\sigma})t^{-n}+\ldots
  \label{eq:w0aroundinfty}
\end{equation}
where now $\bar{c}_{n}(\bar{\sigma})$ is obtained from $c_{n}(\sigma)$
by $\delta_0\leftrightarrow\delta_{3}$ and
$\sigma\rightarrow\bar{\sigma}$.  

In order to understand the relation between the three expansions, let
us take a geometrical view of the parameters. Firstly, let us repeat
that $\{\sigma,\eta\}$ are coordinates for the character manifold of
the four-punctured sphere. With the expansions $\mathscr{W}_t$ and
$\mathscr{W}_u$ we can define two variables
\begin{equation}
  \tilde{\eta}=\frac{1}{\pi i}\frac{\partial \mathscr{W}_t}{\partial \tilde{\sigma}},
  \quad\quad\quad
  \bar{\eta}=\frac{1}{\pi i}\frac{\partial \mathscr{W}_u}{\partial \bar{\sigma}},
\end{equation}
in such a way that two new sets of coordinates are defined, namely
$\{\tilde{\sigma},\tilde{\eta}\}$ and $\{\bar{\sigma},\bar{\eta}\}$.
All three parametrize the same character manifold.

On the other hand, each of the functions $\mathscr{W}_s$,
$\mathscr{W}_t$ and $\mathscr{W}_u$ can be interpreted as generating
functions for the symplectomorphism between the accessory parameter
space of the Heun equation, parametrized by $\{t,c\}$ and the manifold
of monodromy data, labelled by the appropriate pair of coordinates of
the character manifold defined in the preceeding paragraph. As such,
they are linked by the canonical transformations
\begin{equation}
  \{t,\sigma\}\longrightarrow \{1-t,\tilde{\sigma}\},
  \quad\text{and}\quad
  \{1-t,\tilde{\sigma}\}\longrightarrow \{1/t,\bar{\sigma}\},
  \label{eq:symplectictau}
\end{equation}
between $\mathscr{W}_s$ and $\mathscr{W}_t$, and $\mathscr{W}_t$ and 
$\mathscr{W}_u$, respectively. They are in principle
different solutions of the Hamilton-Jacobi equation associated to the
Painlevé VI system. In the complex $t$ plane, we can put conditions on
the monodromy parameters in such a way that the expansions for
$\mathscr{W}$ above have overlapping domains of
convergence\footnote{For the semiclassical limit of conformal blocks
  in unitary CFTs, mutual convergence seems to follow from the usual
  arguments of OPEs between primary operators. Alternatively, one
  could think of the implicit definition given by
  \eqref{eq:c0around0}.}, in these domains, if for the same value of
$t$, the monodromy parameters are equivalent in the sense that they
represent the same point in the character manifold, we can say that
they are equivalent in the sense that their accessory parameter are
the same, after a simple transformation of variables. This fact means
that the canonical transformation linking any two expansions are
functions of the monodromy parameters alone. For instance,
$d\mathscr{W}_s=d\mathscr{W}_t+\pi i (\eta
d\sigma-\tilde{\eta}d\tilde{\sigma})$, with the extra term being the
gradient of a function of the monodromy variables
$\sigma,\tilde{\sigma}$ alone, due to the symplectic structure of the
monodromy manifold itself. Even though the restriction of this
function to the Lagrangian submanifold defined by the zero of the tau
function \eqref{eq:c0tau6} will introduce a complicated $t$ dependence
for the conformal block, we will consider in this case that the three
expansions describe the asymptotic dependence on the $t$ variable of a
single globally defined function $\mathscr{W}$. 

In the regular case, the three patches of the expansion for the
conformal block have been motivated by the relatively simple
transformation law of the Painlevé VI tau function
\eqref{eq:taubraidtrans}, which is in itself due to the $D_4$ symmetry
group of poles of the Heun equation. The symplectic transformations
\eqref{eq:symplectictau} can be thought of as the action of the braid
transformations restricted to the Lagrangian submanifold of
parameters of the tau function defined by the zero condition of
$\tau_{VI}$ in \eqref{eq:c0tau6}. Although this condition does not
have a straightforward CFT interpretation -- at least one that we know
of -- it was checked, both algebraically (up to order $t^6$) and
numerically (to order $t^{64}$) in
\cite{BarraganAmado:2021uyw}. Specifically, we checked that the
derivative of $\mathscr{W}$ defined in the procedure above is the
value of $\eta$ obtained by formally inverting the condition
$\tau_{VI}=0$. As we will see in the following, this property carries
on to confluent limits of $\tau_{VI}$ and similar results can be
extracted for the irregular versions of $\mathscr{W}$.

\section{The irregular semiclassical conformal block}

Confluent limits of the Heun equation are obtained in the
semiclassical limit of irregular conformal blocks
\cite{Gaiotto:2012sf}. Let us define the rank 1 irregular operator
\begin{equation}
  V_{\alpha,\gamma}(z)=\lim_{\Lambda\rightarrow \infty}
  \left(\frac{\gamma}{\Lambda}\right)^{\frac{1}{2}(Q+\alpha_+)(Q+\alpha_-)}
  V_{\alpha_+}(z+\gamma/\Lambda)V_{\alpha_-}(z),
\end{equation}
where $\alpha_\pm=\pm\Lambda+\alpha/2-Q/2$. Starting from this
definition we can find that the state $|\alpha,\gamma\rangle
=V_{\alpha,\gamma}(0)|0\rangle$ belongs to a Whittaker module with 
\begin{equation}
  L_0|\alpha,\gamma\rangle = \left(\Delta(\alpha)+\gamma\frac{\partial}{\partial
      \gamma}\right)|\alpha,\beta\rangle,\qquad
  L_1|\alpha,\gamma\rangle =
  \tfrac{1}{2}\gamma(Q-\alpha)|\alpha,\gamma\rangle,\qquad
  L_2|\alpha,\gamma\rangle = -\tfrac{1}{4}\gamma^2|\alpha,\gamma\rangle,
\end{equation}
and $L_n|\alpha,\gamma\rangle = 0$ for $n>2$. In this expression and
below we will take $\Delta(\alpha)=\frac{1}{4}(Q^2-\alpha^2)$. We note
that, unlike $\alpha$, $\gamma$ is not invariant under global
transformations, so the eigenvalue of the BPZ dual $\langle
\alpha,\gamma|$ under $L_{-2}$ will not be $-\tfrac{1}{4}\gamma^2$.

The analogue of the BPZ identity \eqref{eq:bpzregular} for one
Whittaker operator and $n-1$ primaries is now
\begin{multline}
  \left(\frac{1}{b^2}\frac{\partial^2}{\partial z^2}-
    \left(\frac{1}{z}+\frac{1}{z-1}\right)\frac{\partial}{\partial z}+
    \frac{\Delta_0}{z^2}+\frac{\Delta_{n-2}}{(z-1)^2}+
    \sum_{k=1}^{n-3}\frac{\Delta_k}{(z-z_k)^2}
    +\sum_{k=1}^{n-3}\frac{z_k(z_k-1)}{z(z-1)(z-z_k)}\frac{\partial}{\partial z_k}
    +\right. \\ \left.
    +\frac{\Delta(\alpha)-\Delta_{(1,2)}-\sum_{k=0}^{n-2}\Delta_k}{z(z-1)}
    +\frac{\gamma_0(Q-\alpha)}{2z}-\frac{\gamma_0^2}{4}
    +\frac{\gamma_0}{z(z-1)}\frac{\partial}{\partial \gamma_0}
  \right)\mathscr{G}^c_{n}(z,z_k;\alpha_k,\beta_k,\gamma_0)=0,
  \label{eq:bpzirreg}
\end{multline}
where $\gamma_0$ is proportional to $\gamma$, after using the global
transformations to fix the position of the Whittaker operator at $z=\infty$. 
  
We start by considering $\mathscr{G}^c_{2}$, the irregular analogue of
the conformal block involving one degenerate operator solved in
\cite{Teschner:1995yf}. Its solutions are given in terms of confluent
hypergeometric functions 
\begin{multline}
  \mathscr{G}^c_{2}= (z-1)^{2\Delta_{(1,2)}}\biggl[
  A^c_-z^{\frac{b}{2}(Q+\alpha)} 
  \,{_1F_1}\left(\tfrac{1}{2}b(\alpha_1+\alpha-b); b(Q+\alpha-b);
    b\gamma_0z\right)
   +\\
   A^c_+ z^{\frac{b}{2}(Q-\alpha)}
  \,{_1F_1}\left(\tfrac{1}{2}b(\alpha_1-\alpha-b); b(Q-\alpha-b);
    b\gamma_{0}z\right)\biggr]\exp\left(-\tfrac{1}{2}b\gamma_{0}z\right).
  \label{eq:irregular2ndkind}
\end{multline}
Again, using the OPE for the null field \eqref{eq:nullOPE}, we
find the constants $A^c_{\pm}$ in terms of the two-point functions
involving Whittaker operators. These were constructed in
\cite{Gaiotto:2012sf} and \cite{Bonelli:2022ten} from the DOZZ
formula.

The irregular BPZ equation \eqref{eq:bpzirreg} has a similar
semiclassical limit to \eqref{eq:bpzregular}
\begin{equation}
  \alpha_k=\frac{\theta_k}{b},\qquad \beta_k=\frac{\sigma_k}{b},\qquad
  \gamma_0=\frac{t}{b},\qquad b\rightarrow 0,
\end{equation}
and $\mathscr{G}^c_{n}$ is again expected to exponentiate
\begin{equation}
  \mathscr{G}^c_{n}(z,z_k;\alpha_k,\beta_k,\gamma_0)=
  \psi^c(z,z_k;\alpha_k,\beta_k,\gamma_0)\exp\left[\frac{1}{b^2}
    \mathscr{W}^c(z_k;\theta_k,\sigma_k;t)\right](1+{\cal O}(b^2)).
\end{equation}
For the conformal block with one Whittaker operator
-- $V_{\alpha,\gamma}$ -- and two primary operators -- $V_{\alpha_1}$
and $V_{\alpha_0}$ -- in the semiclassical limit --
$\alpha=\theta_\star/b$ and $\gamma_0=t/b$ -- the BPZ identity
results in the confluent Heun equation \cite{CarneirodaCunha:2019tia}
\begin{equation}
  \left(\frac{\partial^2}{\partial z^2}+
    \frac{\delta_0}{z^2}+\frac{\delta_1}{(z-1)^2}+
    \frac{\delta_\star-\delta_0-\delta_1+tc^c}{z(z-1)}+
    \frac{t(1-\theta_\star)}{2z}-\frac{t^2}{4}\right)
  \psi^c(z,z_k;\theta_k,\sigma_k,t)=0,
  \label{eq:confheun}
\end{equation}
with $\delta_\star=(1-\theta_\star^2)/4$ and 
\begin{equation}
  c^c=\frac{\partial \mathscr{W}^c}{\partial t}=
  -\frac{\partial}{\partial t}\log\tau_{V}(\theta^-_k;\sigma,\eta;t)+
  \frac{\delta_0-\delta_\star}{4t}-\frac{(1-\theta_1)^2}{4t}
  \label{eq:irregconfblktau5}
\end{equation}
where the arguments $\theta^-_k=\{\theta_0,\theta_1-1,\theta_\star\}$
and $\tau_V$ is obtained from the Kyiv formula
\eqref{eq:taunekrasov6} by a confluence limit
\cite{Gamayun:2013auu}
\begin{equation}
  \theta_3=\Lambda+\frac{\theta_\star-1}{2},\qquad
  \theta_2 = -\Lambda +\frac{\theta_\star-1}{2},\qquad
  t\rightarrow\frac{t}{\Lambda},\qquad
  \Lambda\rightarrow \infty,
  \label{eq:confluentlimit}
\end{equation}
and $\sigma$ and $\eta$ are kept fixed. The result is that the
accessory parameter of the confluent Heun equation is given by the
logarithmic derivative of the confluent tau function
\eqref{eq:irregconfblktau5}, with again $\eta$ determined from the
zero locus condition 
\begin{equation}
  \tau_{V}(\theta_k;\sigma+1,\eta;t)=0,\qquad
  \theta_k=\{\theta_0,\theta_1,\theta_\star-1\}.
  \label{eq:irregconfblktoda5}
\end{equation}
Once more the conditions \eqref{eq:irregconfblktau5} and
\eqref{eq:irregconfblktoda5} define the analytical properties of
$\mathscr{W}^c$ from the Painlevé property. Moreover, the small $t$
expansion of $\mathscr{W}^c$ follows the same logic as the
non-confluent case \eqref{eq:w0around0}. First let us make a change of
variables to \eqref{eq:confheun}
\begin{equation}
  z=\frac{z'}{t},\qquad
  \psi^c(z)=z'^{\frac{1}{2}(1-\theta_0)}(z'-t)^{\frac{1}{2}(1-\theta_1)}y(z'),
\end{equation}
which brings the equation to a standard form (as seen in Eqn. (4) in
\cite{daCunha:2021jkm}). We then postulate a Floquet solution of the kind
\begin{equation}
  y(z')=e^{\frac{1}{2}z'}z'^{\frac{1}{2}(\sigma+\theta_0+\theta_1-1)}
  \sum_{n\in\mathbb{Z}}a_nz'^{n},
  \label{eq:floquetconfheun}
\end{equation}
where again $\sigma$ parametrizes the Liouville momentum of the
intermediate channel. Substitution into \eqref{eq:confheun}
yields another three-term recursion relation
\begin{equation}
  A_na_{n-1}-(B_n+t\, C_n)a_n+t\, D_na_{n+1}=0
  \label{eq:recursionsmallt}
\end{equation}
with
\begin{equation}
  \begin{gathered}
    A_n=2\sigma+4n-2\theta_\star, \qquad
    B_n=-(\sigma+2n)^2+\theta_\star^2,\qquad
    C_n= 2\sigma+4n+2\theta_1-2\theta_\star+2-4c^c,\\
    D_n=-(\sigma+2n+\theta_0+\theta_1+1)
    (\sigma+2n-\theta_0+\theta_1+1).
  \end{gathered}
\end{equation}

Going through the same procedure leading to \eqref{eq:c0around0}, we
find the expansion for $c^c$,
\begin{equation}
  c^c(\sigma,t)=c^c_{-1}(\sigma)t^{-1}+c^c_{0}(\sigma)+
  c^c_{1}(\sigma)t+\ldots + c^c_{n}(\sigma)t^n+\ldots
  \label{eq:c0caround0}
\end{equation}
where the first coefficients are 
\begin{gather}
    c^c_{-1}(\sigma)=\delta_\sigma-\delta_\star,
    \qquad
    c^c_{0}(\sigma)=\frac{(1-\theta_\star)(\delta_\sigma-\delta_0+\delta_1)}{
      4\delta_\sigma},\\
    \begin{split}
      c^c_{1}(\sigma)= \frac{\delta_\sigma-\delta_0-\delta_1}{12} &
      -\frac{(\frac{1}{4}(1-\theta_\star)^2-\delta_\sigma^2)
        ((\delta_0-\delta_1)^2-\delta_\sigma^2)}{
        8\delta_\sigma^3}+\\ &
      +\frac{(\delta_\sigma^2-\frac{9}{16}(1-\theta_\star)^2)
        (\delta_\sigma^2+2\delta_\sigma(\delta_0+\delta_1)-
        3(\delta_0-\delta_1)^2)}{
        24\delta_\sigma^2(\delta_\sigma+\frac{3}{4})}.
    \end{split}
\end{gather}
The small $t$ expansion of the semiclassical irregular conformal
block $\mathscr{W}_s^c$ is then defined as the primitive of $c^c(\sigma,t)$:
\begin{equation}
  \mathscr{W}_s^c(\sigma,t)=c_{-1}^c(\sigma)\log t+\Omega^c(\sigma)+
  c^c_{0}(\sigma)t+\frac{1}{2}c^c_{1}(\sigma)t^2+\ldots+
  \frac{1}{n}c^c_{n-1}(\sigma)t^n+\ldots,
  \label{eq:w0csmallt}
\end{equation}
with the $t$-independent term to be determined below.

Because of the confluent limit \eqref{eq:confluentlimit}, however, the 
structure of the expansion of $\mathscr{W}^c$ at large $t$ is quite
different. We can still follow the same argument used for the
Painlevé VI tau function, though: since the tau function in
\eqref{eq:irregconfblktau5} is globally well-defined, we can use the
expansion of the accessory parameter to define $\mathscr{W}^c$
globally.

In parallel to the regular case, the first step is introduce the
$\eta$ parameter
\begin{equation}
  \eta = \frac{1}{2\pi i}\log
  \Theta(\sigma,\theta_1,\theta_0)\Theta^c(\sigma,\theta_\star)+
  \frac{1}{\pi i}\int^tdt'\,c^s(\sigma,t')=
  \frac{1}{\pi i}\frac{\partial \mathscr{W}_s^c}{\partial \sigma},
  \label{eq:etaexptau5}
\end{equation}
where $\Theta$ is as \eqref{eq:theta6}, and
\begin{equation}
  \Theta^c(\sigma,\theta_\star)=
  \frac{\Gamma(1+\sigma)}{\Gamma(1-\sigma)}
  \frac{\Gamma(\tfrac{1}{2}(\theta_\star-\sigma))}{
    \Gamma(\tfrac{1}{2}(\theta_\star+\sigma))},
\end{equation}
which, just as the regular case, can be written in terms of Barnes'
$G$-function. This parameter can be seen to correspond to a zero of
$\tau_V$, \eqref{eq:irregconfblktoda5}, \cite{CarneirodaCunha:2019tia},
and has a similar interpretation as in the non-confluent case
\eqref{eq:etamonodromy}. This sets the $t$-independent term of
\eqref{eq:w0csmallt} as
\begin{equation}
  \Omega^c(\sigma)=\frac{1}{2}\int^\sigma d\sigma'\log
  \Theta(\sigma',\theta_1,\theta_0)\Theta^c(\sigma',\theta_\star).
\end{equation}

As with the regular case, the conditions
\eqref{eq:irregconfblktau5} and \eqref{eq:irregconfblktoda5} have a geometrical
meaning in the sense that they do not depend on the coordinates chosen
to parametrize the monodromy manifold. In order to describe the
quantities relevant for the large $t$ expansion of the conformal block
let us define, following 
\cite{Lisovyy:2018mnj},
\begin{equation}
  X_+=\lim_{\Lambda\rightarrow +i\infty}2e^{\pi
    i\Lambda}\cos\pi\bar{\sigma},\qquad\text{and}\qquad
  X_-=\lim_{\Lambda\rightarrow +i\infty}2e^{\pi
    i\Lambda}\cos\pi\tilde{\sigma}.
\end{equation}
which, when written explicitly in terms of $\sigma$ and $\eta$, find
\begin{multline}
  \sin^2\pi\sigma (X_{\pm}\mp i e^{\pm \pi
    i(\theta_1+\frac{1}{2}\theta_\star)}) = \\
   2\cos\tfrac{\pi}{2}(\sigma-\theta_1+\theta_0)
  \cos\tfrac{\pi}{2}(\sigma-\theta_1-\theta_0)
  \sin\tfrac{\pi}{2}(\sigma-\theta_\star)
  e^{\mp \frac{\pi}{2}i\sigma}(e^{2\pi i\eta}-1)+\\
  +2\cos\tfrac{\pi}{2}(\sigma+\theta_1+\theta_0)
  \cos\tfrac{\pi}{2}(\sigma+\theta_1-\theta_0)
  \sin\tfrac{\pi}{2}(\sigma+\theta_\star)
  e^{\pm\frac{\pi}{2}i\sigma}(e^{-2\pi i\eta}-1).
  \label{eq:xpmsigma}
\end{multline}


We will now define the parameter $\nu$ as $e^{2\pi i\nu}=X_-$, which
will play the role of spectral parameter for the classical conformal
blocks in the $t\rightarrow \infty$ limit. The eigenvalue $c^c(\nu;t)$,
now seen as a function of $\nu$, can be computed by considering the
following expansion for the solution of \eqref{eq:confheun}
\begin{equation}
  y(z')=\sum_{n\in\mathbb{Z}}a_n e^{-\frac{1}{2}z'}z'^{\theta_0}\,F_n(z'),\qquad
  \,F_n(z')={_1F_1}
  (\tfrac{1}{2}\theta_0+\tfrac{1}{4}(1+\theta_\star)-
  \nu-n;1+\theta_0;z').
  \label{eq:infconfansatz}
\end{equation}
This Ansatz comes from the expression of the irregular conformal block
involving the degenerate operator \eqref{eq:irregular2ndkind}. The
$\nu$ parameter represents the shift of the formal monodromy to the 
irregular operator due to the OPE with the primary, as described in
\cite{Lisovyy:2018mnj}. From the analytical perspective, the expansion
parameter $\nu$ is related to the Stokes parameter of the expansion,
pending of course the thorny question of its convergence. From the
connection formula for confluent hypergeometric equation,
\begin{equation}
  {_1F_1}(a;b;z)=e^{-\pi i a}\frac{\Gamma(b)}{\Gamma(b-a)}U(a;b;z)+
  e^{\pm\pi i(b-a)}\frac{\Gamma(b)}{\Gamma(a)}e^{z}U(b-a;b;e^{\pm \pi
    i}z),
\end{equation}
where $U(a;b;z)$ is the Tricomi function, defined as the solution of
the confluent hypergeometric differential equation with asymptotics 
$\propto z^{-a}$ for $z\rightarrow\infty$ in the sector $|\arg
z|<\tfrac{3\pi}{2}$. It is a straightforward exercise to use this
connection formula to show that $\nu$ parametrizes the corresponding
connection formula for the solution of the confluent Heun equation
parametrized by \eqref{eq:infconfansatz}. 

We follow through the BPZ equation
\eqref{eq:confheun}, using the confluent hypergeometric equation to
replace the second derivative of ${_1F_1}$, arriving at 
\begin{multline}
  \sum_{n\in\mathbb{Z}} a_n\bigg[
    (1-\theta_1)\,z\frac{\partial}{\partial z}F_n(z)-
    \left(\nu+n+\frac{1}{2}(1+\theta_1) -
      \frac{1}{4}\theta_\star\right)\,zF_n(z)-\\ 
    -\left(t\left(\nu+n-\frac{1}{4}(1-\theta_\star)-c^c\right)+
      \frac{1}{4}\theta_\star^2-\frac{1}{4}(1+\theta_0-\theta_1)^2
    \right) F_n(z)\bigg]=0.
\end{multline}
using the following properties of Kummer's function
\begin{equation}
  \begin{gathered}
    z\frac{\partial}{\partial z} {_1F_1}(a;b;z) = a\,( {_1F_1}(a+1;b;z)-\,
    {_1F_1}(a;b;z)),\\
    z\,{_1F_1}(a;b;z)=a\,{_1F_1}(a+1;b;z)-(2a-b)\,{_1F_1}(a;b;z)+
    (a-b)\,{_1F_1}(a-1;b;z),
  \end{gathered}
\end{equation}
we arrive at the $3$-term recursion 
\begin{equation}
  \begin{gathered}
    A_n a_{n-1} - (B_n+t\,C_n) a_n + D_n a_{n+1} = 0,\\
    A_n=(2\nu+2n-\theta_1-\tfrac{1}{2}(3-\theta_\star))
    (2\nu+2n+\theta_0-\tfrac{1}{2}(1+\theta_\star)),\\
    B_n=8(\nu+n)^2-\theta_0^2-\theta_1^2+\tfrac{1}{2}(1+\theta_\star)^2,\\
    C_n = -4\nu-4n-4c^c+1-\theta_\star,\\
    D_n =(2\nu+2n+\theta_1+\tfrac{1}{2}(1+\theta_\star))
    (2\nu+2n-\theta_0+\tfrac{1}{2}(3-\theta_\star)).
    \label{eq:recursionM}
  \end{gathered}
\end{equation}
The recursion equation is again solved by continued fractions. Using
\eqref{eq:contfracheun} once more, we obtain the first terms of the
expansion for the accessory parameter as
\begin{equation}
  \begin{gathered}
    \label{eq:c0clarget}
    c^c(\nu,t) = \bar{c}^c_{0}(\nu) +
    \bar{c}^c_{1}(\nu)t^{-1}+\bar{c}^c_{2}(\nu)t^{-2}+
    \bar{c}^c_{3}(\nu) t^{-3}+\ldots,\\
    \bar{c}^c_{0}(\nu)=-\nu+\tfrac{1}{4}(1-\theta_\star),\qquad
    \bar{c}^c_{1}(\nu)=2\nu^2+\delta_0+\delta_1+
    \tfrac{1}{8}(1+\theta_\star)^2-\tfrac{1}{2},\\
    \bar{c}^c_{2}(\nu)=4\nu^3+(2\delta_0+2\delta_1-\tfrac{1}{4}
    (1-\theta_\star)^2)\nu-\tfrac{1}{2}(1-\theta_\star)(\delta_0-\delta_1),\\
    \begin{split}
      \bar{c}^c_{3}(\nu)=20\nu^4+(4+12\delta_0+&12\delta_1-\tfrac{3}{2}
    (1-\theta_\star)^2)\nu^2-4(\delta_0-\delta_1)(1-\theta_\star)\nu-\\ &
    -\tfrac{1}{4}(1-\theta_\star)^2(1-\delta_0-\delta_1)-4\delta_0\delta_1
    +\tfrac{1}{64}(1-\theta_\star)^4,
    \end{split}
  \end{gathered}
\end{equation}
which can be seen to agree with the semiclassical limit of the
confluent conformal block of the second kind presented in
\cite{Lisovyy:2018mnj}.  We can now define the expansion for the
semiclassical confluent conformal block itself 
\begin{equation}
  \mathscr{W}_u^c(\nu;t)=\bar{c}_{0}^c(\nu)t+\bar{\Omega}^c(\nu)+
  \bar{c}_{1}^c(\nu)\log t+\ldots
  -\frac{1}{n}\bar{c}_{n+1}^c(\nu)t^{-n}+\ldots.
  \label{eq:w0clarget}
\end{equation}
For completeness, we can determine the monodromy parameter conjugate
to $\nu$
\begin{equation}
  \rho = -\frac{1}{\pi i}\log
  \bar{\Theta}^c(\nu,\theta_\star,\theta_0)
  \bar{\Theta}^c(\nu,-\theta_\star,\theta_1)+
  \int^tdt'\,c^{s}(\nu,t')=
  \frac{1}{\pi i}
  \frac{\partial \mathscr{W}_u^c}{\partial \nu},
  \label{eq:rhodefinition}
\end{equation}
where
\begin{equation}
  \bar{\Theta}^c(\nu,\theta_\star,\phi)
  =\frac{1}{2\pi i}\Gamma(\tfrac{1}{4}(1-2\phi-\theta_\star+4\nu))
  \Gamma(\tfrac{1}{4}(1+2\phi-\theta_\star+4\nu))
\end{equation}
and $\rho$ is related to monodromy data by $e^{\pi
  i\rho}=1-X_+X_-$. We can verify that $\rho$ computed from
\eqref{eq:rhodefinition} satisfies
\begin{equation}
  \tau_{V}(\theta_k;\nu,\rho;t)=0, \qquad
  \theta_k=\{\theta_0,\theta_1,\theta_\star-1\},
\end{equation}
in the large imaginary $t$ expansion presented in
\cite{Lisovyy:2018mnj}. This allows us to find the $t$-independent
term in \eqref{eq:w0clarget} as
\begin{equation}
  \bar{\Omega}^c(\nu)=\int^\nu d\nu' \log
  \bar{\Theta}^c(\nu',\theta_\star,\theta_0)
  \bar{\Theta}^c(\nu',-\theta_\star,\theta_1),
\end{equation}
which again can be expressed in terms of Barnes' $G$-functions. 
In parallel with the regular case, we then conjecture that the
expansions of $\mathscr{W}^c$ are related by a canonical
transformation in the Lagrangian submanifold of monodromy data
determined by $\tau_V=0$ \eqref{eq:irregconfblktoda5}, parametrized by
$\{t,\sigma\}$ for small $t$ and by $\{1/t,\nu\}$ for large $t$. The
two functions $\mathscr{W}^c_s$ and $\mathscr{W}^c_u$ solve the same
Hamilton-Jacobi equation for the isomonodromic problem and have the
same initial conditions when $\{\sigma,\eta\}$ and $\{\nu,\rho\}$
represent the same point in the character manifold. Furthermore, since
the accessory parameter derived from them coincide in the overlapping
regions of convergence, the canonical transformation relating them is
a function of the monodromy parameters alone.

As a simple test, we recover from \eqref{eq:w0clarget} the accessory parameter
(eigenvalue) expansion for the Mathieu equation by taking
$\theta_0=\theta_1=-1/2$, $\theta_\star=1$, $t=-8h$ and $\nu=-s/4$
in the notation of \href{https://dlmf.nist.gov/28.16}{NIST's} Digital
Library for Mathematical Functions,
\begin{equation}
  -t\bar{c}_{\mathrm{Mathieu}}^c(h)=2sh-\frac{1}{2^3}(s^2+3)-
  \frac{1}{2^7h}(s^3+3s)-\frac{1}{2^{12}h^2}(5s^4+34s^2+9)+
  {\cal O}(h^{-3})
\end{equation}
which can be compared the large $t$ expansion is listed at \S
28.16 through
$\lambda(h)=\frac{1}{4}-2h^2-t\bar{c}_{\mathrm{Mathieu}}^{c}(h)$,
(see also \cite{Gavrylenko:2020gjb}). The extra factor of
$\tfrac{1}{4}-2h^2$ comes about by bringing the Mathieu equation into the
standard Heun format \eqref{eq:confheun}.

A few extra comments about \eqref{eq:w0clarget} are in order. First, the
large $t$ expansion of the Painlevé V tau function has only been
constructed so far along the rays $\arg t = 0, \frac{\pi}{2}$. The
existence of Stokes sectors for the large $t$ expansions of the
confluent hypergeometric functions make us suspect that any expansion
like \eqref{eq:w0clarget} is at best asymptotic. Like the regular
case, convergence can be inferred from OPE arguments in the case
where the parameters can be realized from unitary CFTs. For generic
$\nu$ and $\delta_k$, however, there is no such argument\footnote{We
  thank Referee 1 for studying the expansion and showing that the
  large-order behavior of the coefficients of \eqref{eq:w0clarget}
  display the factorial growth associated to asymptotic series.}.

The asymptotic character of the expansion \eqref{eq:w0clarget}
notwithstanding, we have found, however, numerical evidence that the
implicit definition of $c^c$ provided by \eqref{eq:recursionM} has a finite
radius of validity for generic monodromy parameters. First and
foremost,  remark that  the coefficient of the highest term in $\nu$ is
the $R(\nu)$ polynomial, defined as the formal power series with
constant term $0$ defined by 
\begin{equation}
  \nu=\sum_{n=0}^\infty \frac{1}{n+1}
  \begin{pmatrix} 2n \\ n \end{pmatrix}^2
  R^{n+1}(\nu)= R(\nu)\,{_2F_1}(1/2,1/2;2;16R(\nu)),
\end{equation}
which appears on rooted planar Eulerian orientations, see
\cite{2019arXiv190207369B} for details and an application to the
six-vertex model, where the corresponding conformal block appears as
the partition sum of the system. Given that the elliptic functions
that are usually related to the hypergeometric functions are in fact
analytic, it seems that so is the series \eqref{eq:w0clarget}, at least in this
limit. From the AGT perspective, keeping the highest term in $\nu$ and
$\theta_k$ corresponds to taking the ``Seiberg-Witten'' limit of the
Nekrasov functions, in the semiclassical (Nekrasov-Shatashvili)
limit. In a straightforward exercise, one can show that the first WKB
approximation for the accessory parameter is expressible in terms of
elliptic integrals\footnote{We thank O. Lisovyy for this remark.}. 

Last, but not least, an aspect of resurgence (see \cite{Dunne:2019aqp} for a
review focussed in Painlevé transcendents) is also present here. The
expansion for \eqref{eq:w0clarget} is actually buried in the small
$t$ recursion relation. If one takes \eqref{eq:recursionsmallt}
with a large $t$ expansion for $c^c(\sigma,t)$:
\begin{equation}
  \tilde{c}^c(\sigma,t)=\tilde{c}^c_{-1}(\sigma)t+\tilde{c}^c_{0}(\sigma)+
  \tilde{c}^c_{1}(\sigma)t^{-1}+\ldots +
  \tilde{c}^c_{n}(\sigma)t^{-n}+\ldots,
\end{equation}
one recovers the large $t$ expansion \eqref{eq:c0clarget} by the
substitution
\begin{equation}
  \sigma\rightarrow -2\nu-\theta_1-\tfrac{1}{2}(1-\theta_\star).
\end{equation}
This correspondence between the coefficients
$\tilde{c}^c_{n}(\nu)=\bar{c}^c_{n}(\nu)$ has been verified to
order $t^{-6}$ for generic monodromy parameters\footnote{We thank
  Referee 1 for verifying this relation to order $t^{-150}$.}. 

\section{Examples: Angular Spheroidal Harmonics and Teukolsky radial equation}

It has long been known that linear perturbation theory of the
Kerr black hole yields a system of separable equations of the
confluent Heun type \cite{Chandrasekhar1983}. Eigenvalue problems are
natural to consider from the monodromy properties of the solutions,
see \cite{CarneirodaCunha:2019tia,daCunha:2021jkm} for details in this
context. Actual scattering coefficients can be computed using the
Trieste formula \eqref{eq:connectionheun} and the expressions given
here for $\mathscr{W}$, using the map between the accessory parameters
of the differential equations and the monodromy parameter $\sigma$. We
will, however, defer this analysis to future work, and restrict
ourselves to the monodromy considerations.

\begin{table}[thb]
  \begin{center}    
  \scalebox{0.94}{
    \begin{tabular}{|c|c|c|c|c|c|c|}
    \hline
    $a\omega$&  ${}_{-2}\lambda_{20}$ &
    ${}_{-2}\lambda^{\text{lit.}}_{20}$  & ${}_{-2}\lambda_{21}$    & 
    ${}_{-2}\lambda^{\text{lit.}}_{21}$   & ${}_{-2}\lambda_{22}$    &
    ${}_{-2}\lambda^{\text{lit.}}_{22}$   \\ \hline
    3.0 & $ -1.6055512 $  &  $-1.6135162 $  &
    $  -6.6147648   $ &     $ -6.6144134 $   & $ -11.7605013 $
    &     $ -11.7605089 $    \\ \hline
    4.0 & $  -6.7445626 $  & $ -6.7448527 $  & $ -13.7187995  $
    &   $ -13.7187929 $   & $ -20.8168143  $ &   $ -20.8168144 $
    \\ \hline
    5.0 & $ -13.8102496 $  & $ -13.8102593 $  & $ -22.7797029  $
    &     $ -22.7797028 $   & $-31.8521549 $
    &     $ -31.8521549 $     \\ \hline
    6.0 &  $ -22.8488578 $  & $ -22.8488580 $  & $ -33.8191725  $
    &   $ -33.8191725  $     & $ -44.8762232 $ 
    &     $ -44.8762232 $     \\ \hline
    7.0 & $ -33.8743420 $  & $ -33.8743420 $  & $ -46.8467287 $
    &    $ -46.8467287  $   & $ -59.8936159  $ &     $ -59.8936159 $      \\ \hline
    8.0 & $ -46.8924439 $  & $ -46.8924439 $  & $ -61.8670293 $
    &     $ -61.8670293  $   & $ -76.9067523 $
    &     $ -76.9067523  $     \\ \hline
    9.0 &  $ -61.9059737 $  & $ -61.9059737 $  & $ -78.8825961 $
    &    $ -78.8825961 $    &     $ -95.9170161  $
    &     $ -95.9170161 $     \\ \hline
    10.0 & $ -78.9164728 $  & $ -78.9164728 $  & $  -97.8949078 $
    &    $ -97.8949078  $    & $ -116.9252528 $
    &     $ -116.9252528 $     \\ \hline
    15.0 & $ -193.9463772 $  & $ -193.9463772 $  & $ -222.9310799  $
    &    $ -222.9310799  $   & $ -251.9500731 $
    &     $ -251.9500731 $      \\ \hline
    20.0 & $ -358.9605057 $  & $ -358.9605057 $  & $ -397.9487361 $
    &     $ -397.9487361 $  & $ -436.9625304 $
    &     $ -436.9625304 $    \\ \hline
    \end{tabular}
  }
  \end{center}
  \caption{The comparison between angular eigenvalue for $s=-2$
      $\ell=2$ and $m=0,1,2$ as the value of $a\omega$
      increases. These results were obtained assuming the
      quantization condition \eqref{eq:nuquant}, while the values
      for $\lambda^{\text{lit.}}$ are from
      \cite{BHPToolkit}.} 
  \label{tab:grav}
\end{table}

\subsection{Angular Spheroidal Harmonics}

The angular equation for a frequency $\omega$, azimuthal and
magnetic quantum numbers $\ell,m$, spin $s$ perturbation of a black
hole with angular momentum parameter $a=J/M$ is   
\begin{equation}
\frac{1}{\sin \theta} \frac{d}{d \theta}\left[\sin \theta \frac{d S}{d
    \theta}\right]+\left[a^{2} \omega^{2} \cos ^{2} \theta-2 a \omega
  s \cos \theta-\frac{(m+s \cos \theta)^{2}}{\sin ^{2}
    \theta}+s+{}_{s}\lambda_{\ell,m}\right] S(\theta)=0,
\label{eq:angulareq}
\end{equation}
whose solutions are known as spheroidal harmonics. The associated
problem is to write the allowed values of ${_s\lambda_{\ell,m}}$ that
correspond to regular functions at both North and South poles
$\theta=0,\frac{\pi}{2}$.

The single monodromy parameters can be read by recasting
\eqref{eq:angulareq} in the form \eqref{eq:confheun}
\begin{equation}
  \theta_0 = -(m+s),\qquad \theta_1=m-s,\qquad \theta_\star = 1-2s,
  \qquad t=-4a\omega
  \label{eq:angparameters}
\end{equation}
with accessory parameter
\begin{equation}
  tc^c=-{_s\lambda_{\ell,m}}-2s-2sa\omega-a^2\omega^2.
\end{equation}
The regular singular points at $z=0$ and $z=1$ correspond to the North
and South poles. 

As discussed in \cite{CarneirodaCunha:2019tia}, the condition that
there are solutions regular at both $z=0$ and $z=1$ can be cast in
terms of monodromy data
\begin{equation}
  \sigma_\ell=\theta_0+\theta_1+2\ell+3,\qquad \ell\in\mathbb{N},
\end{equation}
where $2\cos\pi\sigma_\ell = -\Tr \mathbf{M}_1\mathbf{M}_0$. A
straightforward exercise from their definition \eqref{eq:xpmsigma} 
leads us to the values of $X_\pm$: 
\begin{equation}
  X_{\pm} = \pm ie^{\pm\pi i(\theta_1+\tfrac{1}{2}\theta_\star)}
  + 2i\frac{\sin\pi\theta_1\cos\frac{\pi}{2}(\theta_\star+\sigma)}{
    \sin\pi(\theta_1+\theta_0)}e^{\pm\frac{\pi}{2}i\sigma}
  (e^{-2\pi i\eta}-1),
\end{equation}
which simplifies considerably for the parameters
\eqref{eq:angparameters}, given that $s$ and $m$ are half
integers. Considering $2s$ to be an integer before simplifying, and
using the same strategy as the small frequency expansion, 
the $\eta$ parameter can be found to be
\begin{equation}
  \nu_\ell =-\tfrac{1}{2}\theta_1-\tfrac{1}{4}(\theta_\star+1)+\ell
  \label{eq:nuquant}
\end{equation}
with the integer chosen so to reproduce the expansion in
\cite{Casals:2018cgx}. Substituting the terms directly in the
accessory parameter expansion \eqref{eq:c0clarget}, we find
\begin{multline}
  {_{s}\lambda_{\ell,m}}=-a^2\omega^2+4qa\omega+\frac{1}{2}(m^{2}-2s-1)
  -\frac{q^{2}}{2}-\frac{1}{a\omega}\left(\frac{q^3}{8}-\frac{1}{8} 
  q \left(m^2+2 s^2-1\right)-\frac{m
    s^2}{4}\right)\\+\frac{1}{a^2\omega^2}\left(\frac{1}{32} q^2 \left(3
    m^2+6 s^2-5\right)+\frac{1}{64} \left(4 m^2 s^2-m^4+2 m^2+4
    s^2-1\right)+\frac{1}{4} m q s^2-\frac{5
    q^4}{64}\right)\\-\frac{1}{a^3\omega^3}\bigg(\frac{1}{128} m s^2
\left(m^2+2 s^2-13\right)+\frac{1}{512} \left(-36 m^2 s^2+13 m^4-50
  m^2+8s^{4}-100s^{2}+37\right)q\\
-\frac{33}{128}m
s^{2}q^{2}-\frac{1}{256}\left(23 m^2+46 s^2-57\right)q^3+\frac{33
  q^5}{512} \bigg)+\mathcal{O}(a^{-4}\omega^4),
\end{multline}
where $q=2(\ell+s)-m+1$.

Numerically, the expansion \eqref{eq:c0caround0} is also consistent
with the results in the literature \cite{BHPToolkit}, as can be
checked in Table \ref{tab:grav}. We remark that this is actually a
test of the irregular conformal block expansion at large negative
values of $t$, with good agreement ($6$ significant digits) obtained
for $t\approx -20$ and below. 

\subsection{The radial equation}

The radial equation for linear perturbations is
\begin{equation}
\Delta^{-s}\frac{d}{dr}\left(\Delta^{s+1}\frac{dR(r)}{dr}\right)+
\left(\frac{K^{2}(r)-2is(r-M)K(r)}{\Delta}+4is\omega
r-{_s\lambda_{\ell,m}}-a^2\omega^2+2am\omega
\right)R(r)=0,
\label{eq:radialeqn}
\end{equation}
where
\begin{equation}
K(r)=(r^2+a^2)\omega-am,\quad\quad \Delta =
r^2-2Mr+a^2=(r-r_+)(r-r_-).
\end{equation}

The equation can be brought to the canonical form \eqref{eq:confheun},
with parameters
\begin{equation}
  \begin{gathered}
\theta_{0}= s -
i \frac{\omega-m\Omega_{-}}{2\pi T_-}, \qquad
\theta_{1}=  s +
i \frac{\omega-m\Omega_{+}}{2\pi T_+},\qquad
\theta_{\star}=1+2s-4iM\omega,\\
2\pi T_{\pm} = \frac{r_+-r_-}{4Mr_{\pm}}, \qquad
\Omega_{\pm} = \frac{a}{2Mr_{\pm}},
\label{eq:radparameters}
\end{gathered}
\end{equation}
where $r_\pm$ are the radial positions of the inner and outer event
horizons, corresponding to the regular singularities of the
differential equation. 
\begin{equation}
  \begin{gathered}
    t = 2i(r_{+}-r_{-})\omega,\\
    tc^{c}= - {}_{s}\lambda_{\ell,m}-2s +  is(r_+-r_-)\omega +
    2 i(1-2s) M \omega - r_-r_+\omega^2 + 2(r_{+}-r_{-})M\omega^2
    + 4 M^2\omega^2.
%
\label{eq:accparamrad}
\end{gathered}
\end{equation}

\begin{table}[htb]
  \begin{center}
    \begin{tabular}{|c|c|c|}
    \hline
    $n$ &  $M{}_{-2}\omega_{20}$ -- large $t$
       & $M{}_{-2}\omega_{20}$ -- small $t$ \\ \hline
    1 & $0.37367168441804 - 0.08896231568894i$
      & $0.37367168441804 - 0.08896231568894i$ \\ \hline
    2 & $0.34671099687916 - 0.27391487529123i$
      & $0.34671099687916 - 0.27391487529123i$ \\ \hline
    3 & $0.30105345461236 - 0.47827698322307i$
      & $0.30105345461236 - 0.47827698322307i$ \\ \hline
    4 & $0.25150496218559 - 0.70514820243349i$
      & $0.25150496218559 - 0.70514820243349i$ \\ \hline
    5 & $0.20751457981306 - 0.94684489086635i$
      & $0.20751457981306 - 0.94684489086635i$ \\ \hline
    6 & $0.16929940309304 - 1.19560805413585i$
      & $0.16929940309304 - 1.19560805413585i$ \\ \hline
    7 & $0.13325234024519 - 1.44791062616204i$
      & $0.13325234024519 - 1.44791062616204i$ \\ \hline
    8 & $0.09282233367020 - 1.70384117220614i$
      & $0.09282233367020 - 1.70384117220614i$ \\ \hline
    9 & $0.06326350512560 - 2.30264476515854i$
      & $0.06326350512560 - 2.30264476515854i$ \\ \hline
    10& $0.07655346288598 - 2.56082661738151i$
      & $0.07655346288598 - 2.56082661738151i$ \\ \hline
    \end{tabular}
  \end{center}
  \caption{First overtones for $s=-2$ $\ell=2$ in the Schwarzschild black
    hole. To the left, we list the values for $M{}_s\omega_{\ell,m}$ found from
    \eqref{eq:w0clarget}, using the quantization condition
    \eqref{eq:nuradquantcond}. To the right, we present the values
    obtained from the small $t$ expansion of $\mathscr{W}^c$
    \eqref{eq:w0csmallt} with the
    quantization condition \eqref{eq:quantcond}.} 
  \label{tab:gravrad}
\end{table}

The quasi-normal modes (QNM) problem for \eqref{eq:radialeqn} poses a
suitable test for the convergence of the formulas \eqref{eq:w0csmallt} and
\eqref{eq:w0clarget}, since generically $\omega$ -- and thus $t$ --
are complex numbers. Given that there are no natural small parameter
for the problem, actual QNM frequencies of interest are at finite,
specific values of $t$, so the survey has to be done
numerically. One notes that low-lying modes and/or the near  
extremal case $r_-\rightarrow r_+$ can be suitably studied with the
small $t$ expansion \cite{daCunha:2021jkm}. The large $t$
expansion \eqref{eq:c0caround0} will, on the other hand, be useful for
higher excited modes, as well as small rotation parameter. One notes
that for the actual solutions, $t$ in \eqref{eq:accparamrad} lie in
the third quadrant $0 < \arg t < \frac{1}{2}\pi$, where the
expansion of $\tau_{V}$ may not have either the forms given in
\cite{Lisovyy:2018mnj}.

\begin{figure}[htb]
\begin{center}
  \includegraphics[width=0.55\textwidth]{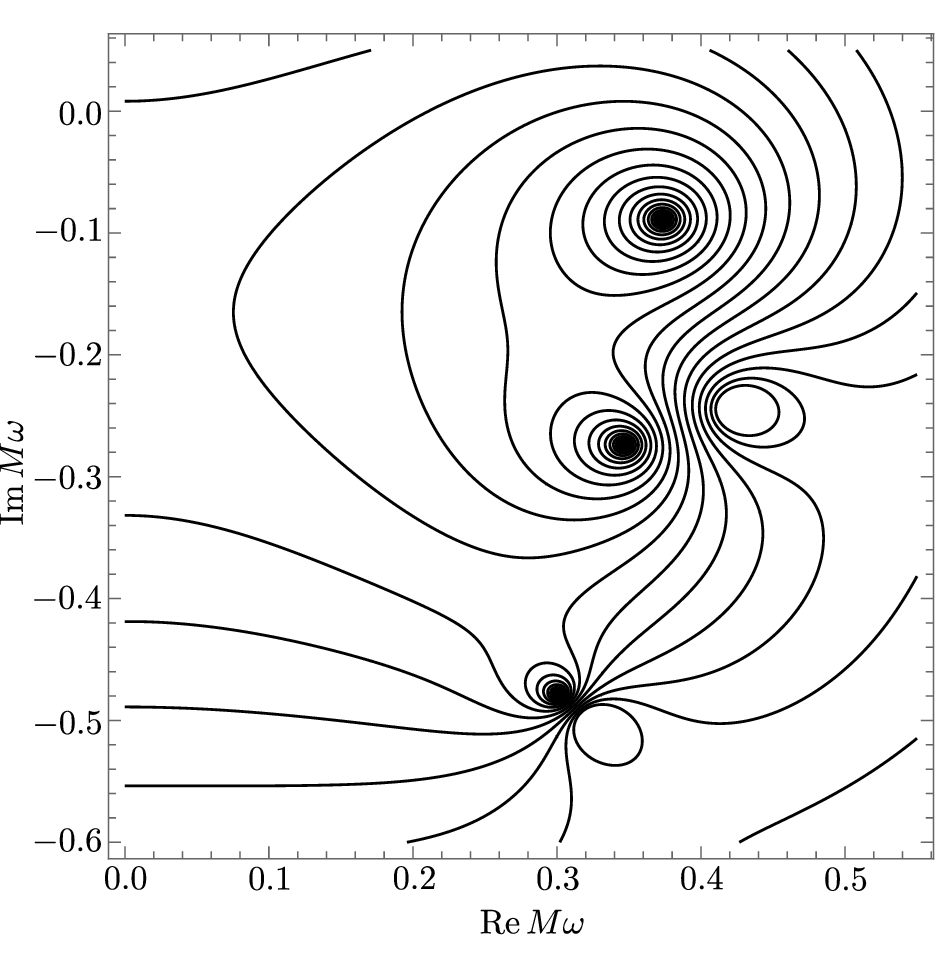}
  \caption{Contour plot for equation \eqref{eq:contfracheun} the
    Schwarzschild $a=0$ case for the radial equation in $M\omega$
    complex plane. Note the first three zeros corresponding to the
    first three overtones.} 
  \label{fig:contour_schw}
\end{center}
\end{figure}

As before, the first step is to translate the boundary conditions to
monodromy data. For the parameters $\sigma$, $\eta$, this was found to
be \cite{CarneirodaCunha:2019tia}
\begin{equation}
    e^{2\pi i\eta}=e^{-\pi  i\sigma}
  \frac{\cos\frac{\pi}{2}(\sigma+\theta_1+\theta_0)
  \cos\frac{\pi}{2}(\sigma+\theta_1-\theta_0)
  \sin\frac{\pi}{2}(\sigma+\theta_*)}{
  \cos\frac{\pi}{2}(\sigma-\theta_1+\theta_0) 
  \cos\frac{\pi}{2}(\sigma-\theta_1-\theta_0)
  \sin\frac{\pi}{2}(\sigma-\theta_*)}.
  \label{eq:quantcond}
\end{equation}
and another straightforward exercise using \eqref{eq:xpmsigma} shows
that the value of $\nu$ is determined:
\begin{equation}
  \nu_n = -\tfrac{1}{2}\theta_1-\tfrac{1}{4}(\theta_\star+1)+
  n,\qquad n\in\mathbb{Z},
  \label{eq:nuradquantcond}
\end{equation}
which now is independent of $\rho$, regardless of the values of
the $\theta_k$. The numerical calculation for the first ten overtones
for the Schwarzschild case $a/M=0$ can be checked in Table
\ref{tab:gravrad},  and match those obtained in the small $t$
expansion obtained previously \eqref{eq:w0csmallt} and the values
accepted in the literature from \cite{Berti:2009kk}. We present the
contour plot for \eqref{eq:contfracheun} in the large $t$ limit with
the radial parameters in  Fig. \ref{fig:contour_schw}, where we can
see that the zeros are simple, a consequence of the Painlevé property. 

In order to compute the values, we used a simple root finding
algorithm with the implicit relation between the accessory parameter
and $t$. We truncate \eqref{eq:contfracheun} to the $N_c$-th level of
recursion -- the $N_c$-th ``approximant'' -- for $A_n,\ldots,D_n$
parameters given by \eqref{eq:recursionM}, using the monodromy
parameters \eqref{eq:radparameters} and the quantization condition
\eqref{eq:nuradquantcond}. As an indication of the convergence of the
procedure, we give the eigenfrequency computed as a function of $N_c$ in
Table \ref{tab:approxNc}.

\begin{table}[htb]
  \begin{center}
  \begin{tabular}{|c|c|c|}
    \hline
    $N_c$&  $M{}_{-2}\omega_{20}$ -- $n=1$ & $M{}_{-2}\omega_{20}$ -- $n=2$   \\
    \hline
    5 &  $\underline{0.373}349851668867-\underline{0.08}95702232125385i$ &
            $\underline{0.34}9043220809013-\underline{0.27}6903141133067i$
    \\ \hline
    10 &  $ \underline{0.3736}99513640235 -
       \underline{0.0889}573826390044i  $ & $
                                            \underline{0.346}574173402192
                                            -
                                            \underline{0.273}507489007237i
                                            $   \\
    \hline
    25 & $ \underline{0.373671}747369450 -
       \underline{0.0889623}297885258i $ & $
                                           \underline{0.3467}06054355762
                                           -
                                           \underline{0.27391}3724423707i
                                           $   \\
    \hline
    50 & $ \underline{0.3736716844}70966 -
        \underline{0.0889623156}216435i $ & $
                                            \underline{0.34671}1028512825
                                            -
                                            \underline{0.2739148}94021952i
                                            $   \\
    \hline
    75 & $ \underline{0.373671684418}541 -
        \underline{0.088962315688}6536i $ & $
                                            \underline{0.34671099}7667890
                                            -
                                            \underline{0.27391487}4928415i
                                            $   \\
    \hline
    100 & $ \underline{0.37367168441804}8 -
         \underline{0.0889623156889}415i $ & $
                                             \underline{0.346710996}914372
                                             -
                                             \underline{0.2739148752}78545i
                                             $    \\
    \hline
    125 & $ \underline{0.373671684418042} -
         \underline{0.088962315688935}8i $ & $
                                             \underline{0.3467109968}81306
                                             -
                                             \underline{0.27391487529}2230i
                                             $    \\
    \hline
    ---  & $ 0.373671684418042 - 0.0889623156889357i $ & $
                                                       0.346710996879163
                                                       -
                                                       0.273914875291235i
                                                       $     \\
    \hline
  \end{tabular}
  \end{center}
  \caption{First and second overtones for $s=-2$ $\ell=2$ in Schwarzschild black
    hole found from the implicit relation \eqref{eq:contfracheun} as a
    function of the approximant $N_c$. In the last line we display the
    common digits of the largest $N_c$ obtained from the small and
    large $t$ formulas. In the lines above, we underline the digits in
    common with the value in the last line.} 
  \label{tab:approxNc}
\end{table}

\section{Discussion}

Conformal blocks are special functions important to the
characterization of conformal field theories. They are also relevant
in a variety of applications, of which we have focussed in the
accessory parameter of Fuchsian differential equations in the complex
plane and their confluent limits. In this paper we presented a method of
relating the conformal block expansions at different branch points
which relies on the monodromy information. We gave an explicit
construction in the regular and rank one irregular cases. We
illustrate the construction by recovering known large frequency
expansions for spheroidal harmonics and showing that both expansions
agree numerically on the overtone spectrum for the Schwarzschild black
hole. 

Our goal was to help characterize conformal blocks in the whole of
complex plane for the $t$ variable. Monodromy parameters manifest
the symmetry of the expansions in the regular case, and help
identifying the suitable variables of the coefficients in the
irregular case. We have found that using the symplectic structure to
that end makes for a clearer formulation, at the expense of computing
an extra monodromy parameter -- such as $\eta$ in the small $t$
expansion. More direct methods have been proposed
\cite{Lisovyy:2022flm}, where one can see that $\mathscr{W}$ and
$\mathscr{W}^c$ have a rather complicated singularity structure for
large $t$. In the relations given here, these branch points and poles
can be seen to arise from the transformation between the monodromy
parameters -- for instance between $\sigma,\eta$, which are suitable
near $t=0$, to $\bar{\sigma},\bar{\eta}$ which are suitable for
$t=\infty$ expansions of $\mathscr{W}$. It would be interesting to
investigate whether the singularity structure can be completely
described from the relations proposed here. A related but important
problem is to use these expansions to efficiently compute the
Nekrasov-Shatashvili limit of the Nekrasov functions entering the
expansion of the semiclassical conformal blocks.

The fact that this extra monodromy parameter, such as $\eta$ in
\eqref{eq:etaasw6}, can also be computed from the semiclassical
conformal block expansions may have interesting repercussions for the
isomonodromic tau functions themselves. We can state one simple such
repercussion by noting that $\eta$ thus defined as a function of
$\sigma$, $t$ and the $\theta_k$ correspond to a zero of $\tau_{VI}$
as can be seen in, for instance, \eqref{eq:c0tau6}. Due to the
structure of the expansion of the tau function, it seems natural to
define $\eta$ by formally inverting the Kyiv formula, and thus $t$ and
$\sigma$ are natural parameters for the Lagrangian submanifold of
monodromy parameters, at least for small $t$. When the inversion is
performed, we make use of the quasi-periodicity $\tau(\sigma+2n;t)\propto
\tau(\sigma;t)$ of the tau function and choose a branch of $\sigma$,
where the proportionality factor is independent of $t$. One can in
principle then define an infinite number of monodromy parameters
$\eta_n=\eta(\sigma+2n;t)$ corresponding to multiple branches of
$\sigma$. Each of these $\eta_n$ correspond to a different 
zero of the tau function. Furthermore, we can construct similar zeros
from the quasi-periodicity with respect to $\tilde{\sigma}$ and
$\bar{\sigma}$. The natural question to be considered is whether
all zeros of the isomonodromic tau function can be obtained from this
procedure. 

In the application front, monodromy variables are also
relevant parameters for a number of properties of solutions of the
Heun equation as well as their confluent limits
\cite{Bonelli:2022ten}. Expressions for black hole properties such as
greybody factors and Love numbers have also a clean definition in
terms of $\sigma$ \cite{Bonelli:2021uvf,Consoli:2022eey}, so a natural
question that arises is whether $\nu$ plays a similar role for high
frequency -- which translates to large $t$. More abstractly, we can
transpose the construction to irregular conformal blocks involved in
the Painlevé III transcendent to shed light on the work of
\cite{Gavrylenko:2020gjb}. At any rate, the
applications of conformal blocks to the accessory parameters display a
variety of relevant physical phenomena which can hopefully shed
light on the global structure of these important special functions,
including the generic number of points and higher genus cases.  

\section*{Acknowledgements}

We thank A. Grassi, A. P. Balachandran, A. R. Queiroz, F. Novaes,
A. Lima, M. Lima and specially O. Lisovyy for stimulating discussions
and suggestions. We thank the referees for substancial contribution
to the interpretation of the results and for the help in clarifying
some of the passages. JPC acknowledges partial support from CNPq.


\providecommand{\href}[2]{#2}\begingroup\raggedright\endgroup

\end{document}